\documentclass[graybox]{svmult}

\usepackage{mathptmx}       
\usepackage{helvet}         
\usepackage{courier}        
\usepackage{type1cm}        
\usepackage{makeidx}         
\usepackage{graphicx}        
\usepackage{multicol}        
\usepackage[bottom]{footmisc}
\usepackage{caption}

\makeindex             
                       
\begin{document}

\title*{Phase transitions in biological membranes}
\author{Thomas Heimburg, \today}
\institute{Thomas Heimburg \at Niels Bohr Institute, University of Copenhagen, Blegdamsvej 17, 2100 Copenhagen, Denmark, \email{theimbu@nbi.ku.dk}
}
%
%
\maketitle

\abstract{Native membranes of biological cells display melting transitions of their lipids at a temperature of 10-20 degrees below body temperature. Such transitions can be observed in various bacterial cells, in nerves, in cancer cells, but also in lung surfactant. It seems as if the presence of transitions slightly below physiological temperature is a generic property of most cells. They are important because they influence many physical properties of the membranes. At the transition temperature, membranes display a larger permeability that is accompanied by ion-channel-like phenomena even in the complete absence of proteins. Membranes are softer, which implies that phenomena such as endocytosis and exocytosis are facilitated. Mechanical signal propagation phenomena related to nerve pulses are strongly enhanced. 
The position of transitions can be affected by changes in temperature, pressure, pH and salt concentration or by the presence of anesthetics. Thus, even at physiological temperature, these transitions are of relevance. There position and thereby the physical properties of the membrane can be controlled by changes in the intensive thermodynamic variables. Here, we review some of the experimental findings and the thermodynamics that describes the control of the membrane function.}

\section{Introduction}
\label{sec:Introduction}
It is a remarkable fact that most biomolecules display order transitions close to physiological conditions. DNA unfolds in a temperature regime above 65$^\circ$C depending on the fraction of GC pairs, salt concentration and pH \cite{Marmur1962, Manning1978}, and on the binding of proteins. 
Protein unfolding temperatures are also found above body temperature and range from slightly above physiological temperature to above 100$^\circ$C. At low temperatures, one finds the interesting phenomenon of cold unfolding \cite{Privalov1990}, which has been postulated being the cause for the inactivation of many enzymes at low temperature.  If physiological temperature is assumed to be around 310 K, the distance to the temperature of unfolding is small on the absolute temperature scale (e.g., typically only a few percent). Therefore, transitions can be induced by relatively small changes in transition enthalpy. As a consequence, proteins can be denatured by pressure \cite{Ravindra2003}, by changes in pH, by addition of small hydrogen-bonding molecules such as urea, or by voltage gradients. Thus, the temperature of denaturation of both DNA and proteins depends on the thermodynamic variables temperature, pressure, and the chemical potentials of protons, ions and small molecules - and the electrostatic potential. It is clear that both DNA and protein function depends on variations of these variables and that the state of these molecules can be controlled by changing those variables from the outside. 

Lipid membranes share many of the above features. They display melting of the lipid chains that depends on lipid composition, temperature, hydrostatic and lateral pressure, on ion concentration, pH and the binding or insertion of ligands like proteins or anesthetic drugs. It is surprisingly unknown that also biological membranes display melting transitions at temperatures of about 10-15$^\circ$ below body temperature \cite{Heimburg2005c, Wang2018}. In contrast to proteins, membranes are quasi two-dimensional and macroscopic. They can support phenomena on scale that are larger than individual molecules. The transitions serve as a functional switch that can be controlled sensitively from the outside. 

The membrane transitions and their important functional implications are the topic of this review.

\section{Membrane melting}
\label{sec:Membrane Melting}
\begin{figure}[tb!]
	\includegraphics[width=11.5cm]{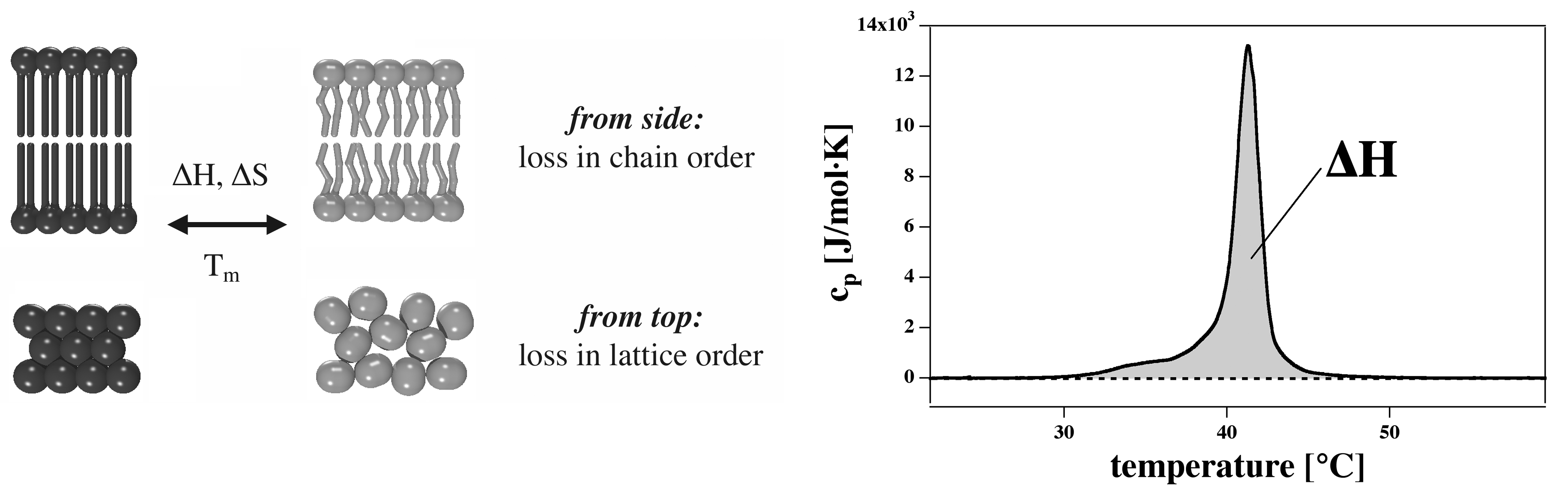}
	\caption{Lipid membrane melting. Left: Schematic drawing of the changes in chain and lattice oder in the melting transition of lipids. Right:Heat capacity profile of the artificial lipid dipalmitoyl phosphatidylcholine (DPPC). The melting transition can be found at 42\,$^\circ$C. The grey-shaded area indicates the melting enthalpy, i.e., the heat of melting.  From \cite{Heimburg2007a}.}
	\label{Figure_DPPC_LUV}       
\end{figure}

\paragraph{\textbf{Artificial membranes}}\label{subsec:artificial membranes}
In the lipid melting transition, membranes change their enthalpy $H$ and entropy $S$, their volume $V$, their area $A$ and their thickness $D$ \cite{Heimburg1998}. The most prominent change on the molecular scale is the loss of order in the lipid hydrocarbon chains. It is mostly due to rotations around the C-C bonds of the chains \cite{Heimburg2007a}. Furthermore, one finds a solid-liquid transition of the crystalline lipid arrangement within the membrane plane, i.e., the lipids can arrange on lattices at low temperature and have more random arrangements at high temperature, with important implications for the solubility of other molecules in the membrane. Consequently, the transitions are denoted as solid-ordered to liquid-disordered transitions (or more trivially as transitions from `gel' and 'fluid'). Fig. \ref{Figure_DPPC_LUV} shows the change in order, and the associated heat capacity (c$_p$-) profile of dipalmitoyl phosphatidylcholine (DPPC) - one of the most frequently studied synthetic lipids. The area under the heat capacity profile is the heat of melting, $\Delta H$. The melting entropy can be deduced from these data because $\Delta S=\Delta H/T_m$ \cite{Heimburg2007a}.  Functions such as the heat capacity are called susceptibilities. They indicate how sensitive a property of the membrane reacts to changes in an intensive variable (such as temperature, pressure, pH or voltage). Beyond the heat capacity, other examples for susceptibilities are  volume and area compressibility \cite{Heimburg1998}, and capacitance \cite{Heimburg2012}. They generally display maxima in the transition, which implies that one can most sensitively adjust the state of the membrane by subtle changes in experimental conditions. It is unlikely that biology does not make use of such an important feature.
The transition temperature of pure lipids depends on chain length and chain saturation. Unsaturated chains display lower melting points. Therefore, the fraction of unsaturated chains provides an estimate for the transition temperature.  Another factor for the melting behavior is the headgroup. Of the uncharged lipid headgroups, phosphatidylethanolamines display higher melting temperatures as phosphatidylcholines. There exist also negatively charged lipid headgroups such as phosphatidylserine, phosphatidylglycerol or phosphatidylinositol. This renders transition temperatures sensitive to surface-active enzymes such as phospholipase $A_1$, $A_2$ \cite{Mouritsen2006, Gudmand2009}, phospholipases C and D, which alter the chain or headgroup structure of the lipids. The melting temperatures of charged lipids display a strong dependence on pH and divalent cations such as calcium, discussed below. The more charged a membrane is, the lower its melting temperature. Consequently, lowering of pH \cite{Trauble1976}, increase in calcium concentration and the binding of basic proteins \cite{Heimburg1996a} increases the melting temperature of biomembranes.

\begin{figure}[tb!]
	\includegraphics[width=11.5cm]{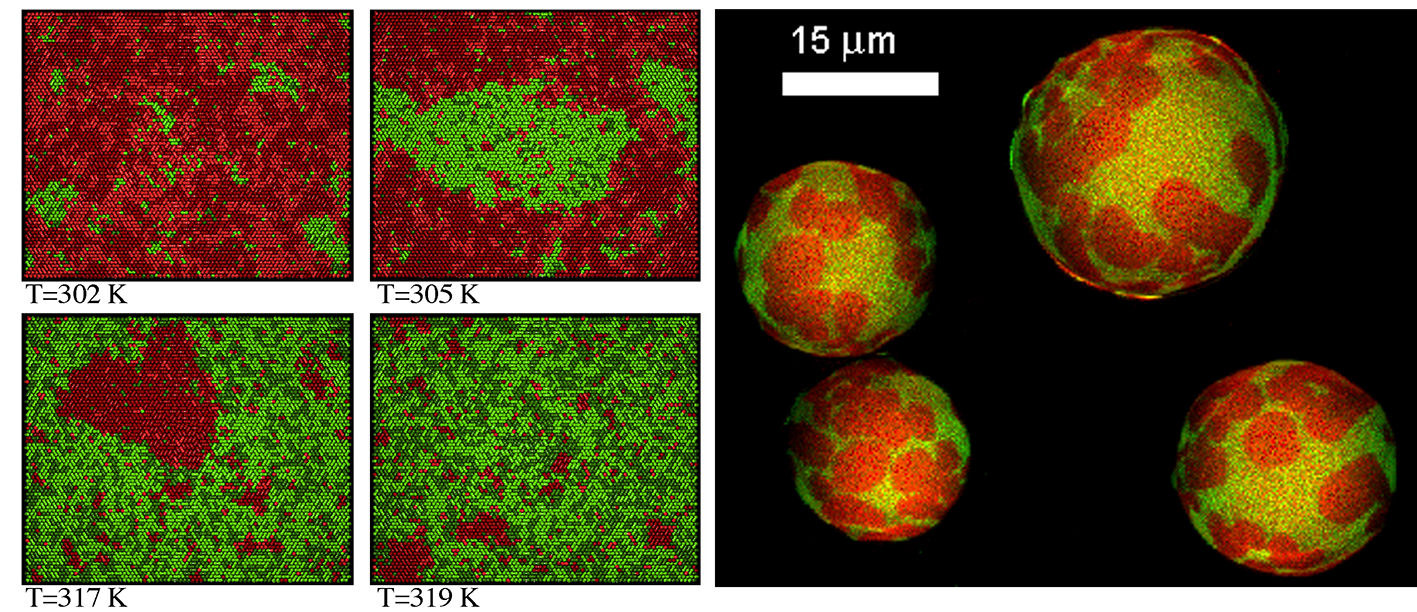}
	\caption{Domain formation in lipid mixtures. Left: Monte-Carlo simulation at four different temperatures. Right: Domain formation in vesicles of a lipid mixture in the transition range as measured by confocal fluorescence microscopy. In both panels, red represents solid domains and green represents liquid domains. From \cite{Hac2005}.}
	\label{Figure_Domains}       
\end{figure}

In the transition, lipid mixtures can phase-separate into phases or domains of different composition and properties \cite{Lee1977}. This is shown in Fig. \ref{Figure_Domains}.  It is a plausible assumption that the frequently discussed `rafts' \cite{Simons1997} in biological membranes (often considered as functional platforms) are just a consequence of domain formation following the rules of phase diagrams \cite{London2002, Almeida2014}. 
Further, as will be shown below, the elastic constants of lipid membranes are especially high in the transition. Within the transition range, membranes are soft, easily compressible, and very flexible (Section \ref{sec:Elastic}). They also display a maximum in permeability (Section \ref{sec:Permeability}). Therefore, the general assumption that lipid membranes are insulators for ions is incorrect close to transitions. The properties of the membranes are also obviously dependent on variables such as temperature, pressure, pH, calcium concentration, and voltage. Therefore, the function of the lipid membrane can be fine-tuned by adjusting these parameters.

\paragraph{\textbf{Biological membranes}}
\label{subsec:biological membranes}
It is much less known that biological membranes also display transitions. There exists quite some literature from the 1970s when the interest in membrane thermodynamics was very high. After the finding of channel proteins and receptors, the interest shifted away from membrane function to more protein-based mechanisms. This led to a very unfortunate neglect of the very important role of the lipid membrane itself. 

On average, biological membrane contain about 50\% proteins and 50\% lipids by mass. This includes the extra-membranous parts of the proteins. Therefore, in the membrane plane itself the lipids outweigh the proteins. There exist some exceptions with extremely high protein content: the purple membrane is effectively a 2D-crystal of bacteriorhodopsin with a lipid:protein ratio of about 1:4. But even in the purple membrane, lipid transitions have been described \cite{Jackson1978}. In contrast, myelin membranes and lung surfactant display low lipid:protein ratios of about 4:1.  For lung surfactant, melting transitions are pronounced and well described \cite{Ebel2001}. The myelin is discussed below.
\begin{figure}[htb]
	\sidecaption[t]
	\includegraphics[width=6cm]{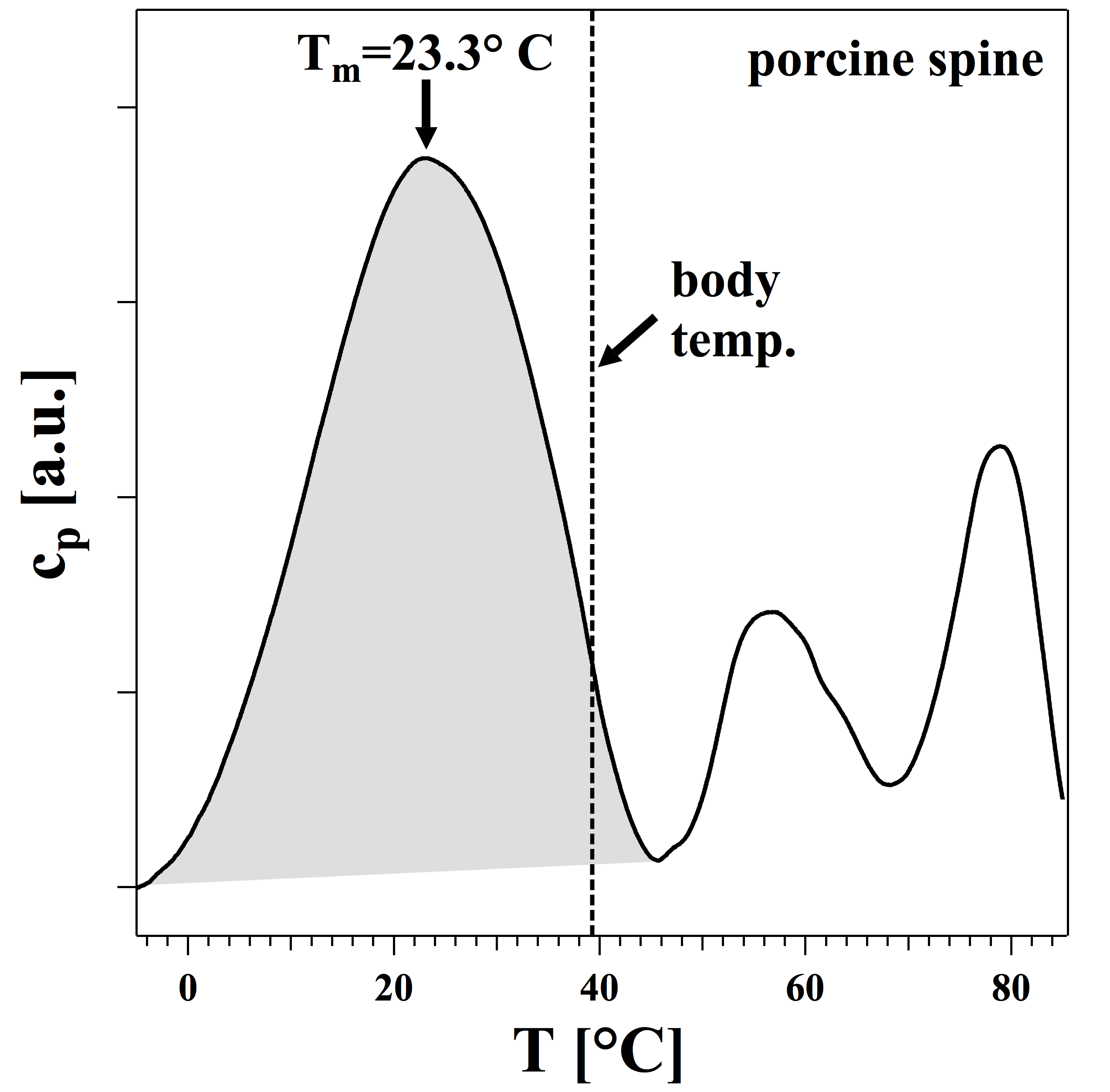}
	\caption{Heat capacity profile of membranes from the spinal cord of pigs. The grey-shaded peak represents lipid melting. The dashed line represents the body temperature of pigs ($\sim$39.3\,$^\circ$C). The peaks above body temperature represent protein unfolding. Adapted from \cite{Wang2018}.}
	\label{Figure_PorcineSpine}       
\end{figure}

Most biomembranes (including all their membrane proteins) display transitions around 10-15$^\circ$ below body temperature. Fig. \ref{Figure_PorcineSpine} shows the melting profile of porcine spine membranes (adapted from \cite{Wang2018}). One can recognize a pronounce maximum at 23.3$^\circ$C (grey-shaded peak) corresponding to lipid melting, and two more maxima corresponding to protein unfolding at 56.5$^\circ$C and 79$^\circ$C, respectively. The latter peaks disappear in a second heating scan because the heat unfolding of proteins is not reversible. One can distinguish the lipid melting from protein unfolding peaks because \cite{Muzic2018}
\begin{itemize}

	\item in contrast to the lipid melting peaks, the protein unfolding peaks are not reversible and disappear in a second heating scan.
	
	\item lipid and protein unfolding peaks display very different pressure dependence.
	
	\item lipid extracts from the native membranes (in the absence of proteins) display transitions in the proximity of the lipid peak found in the recordings of native membranes.
\end{itemize} 

Similar melting profiles have been reported for lung surfactant \cite{Ebel2001, Heimburg2005c, Wang2018}, \textit{E.coli} membranes \cite{Heimburg2005c}, \textit{bacillus subtilis} membranes \cite{Heimburg2005c} and various species of cancer cells \cite{Hojholt2018}. Further, such transitions have been found in native membranes from chicken spine, rat spine and goat spine (unpublished data from our lab). In all of these preparations, the lipid melting peak was found about 7-15 degrees below body temperature. 

The assumption that transitions in biomembranes represent a meaningful feature of practically all cells that is actively maintained by the cell will guide the below considerations.

\paragraph{\textbf{Adaptation}}
\label{subsec:adaptation}

\begin{figure}[t]
	\includegraphics[width=11.5cm]{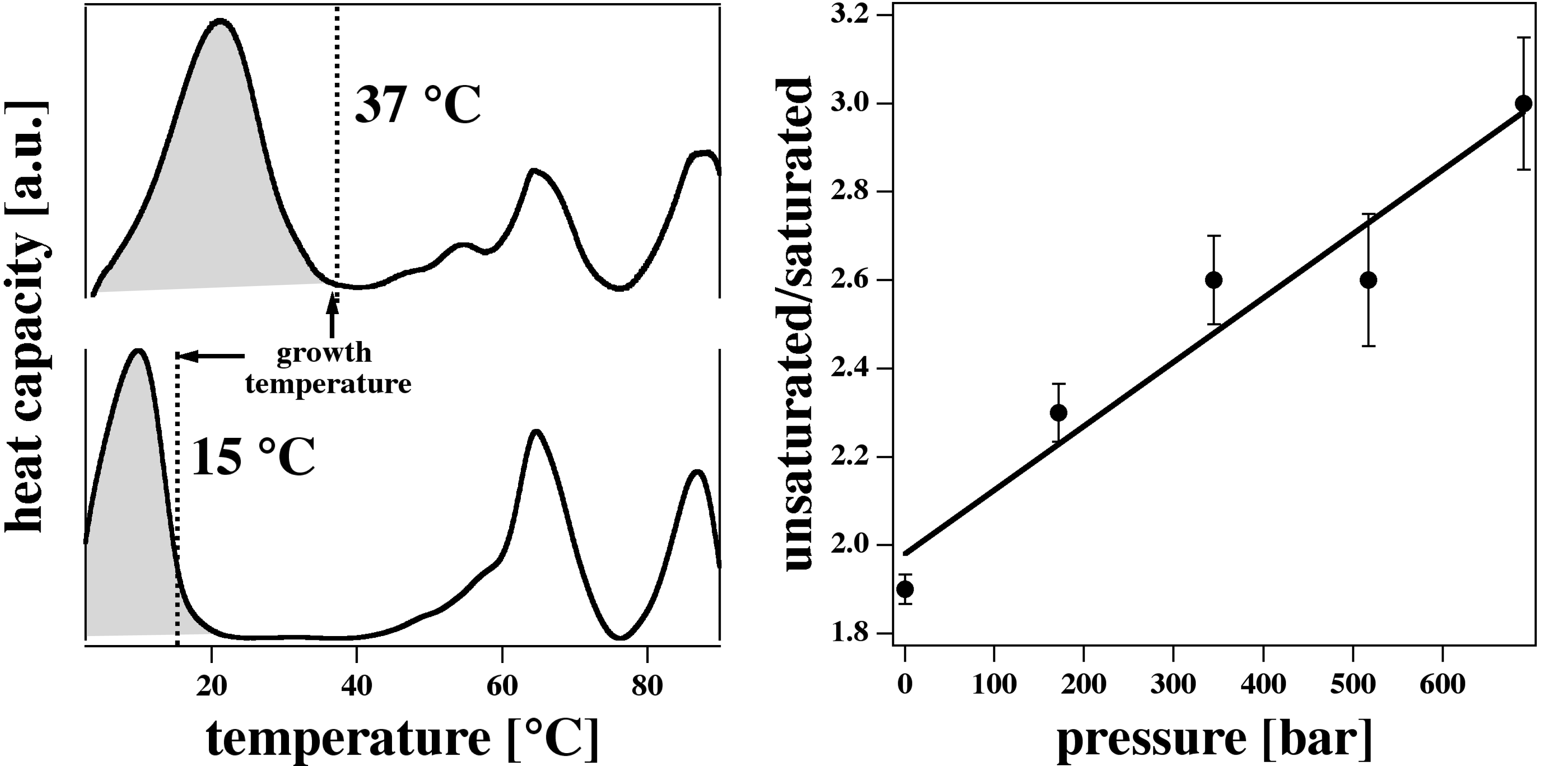}
	\caption{Adaptation. Left: Heat capacity profiles of \textit{E. coli} membranes grown at 37$^\circ$C (top) and 15$^\circ$C (bottom). The dashed lines represent growth temperature. The lipid melting peak (grey-shaded) adapt to growth temperature. Protein unfolding peaks remain unaltered. From \cite{Heimburg2007a}. Right: Ratio of lipids with low and high melting temperatures (unsaturated versus saturated) as a function of pressure of the barophilic deep sea bacterium CNPT3. Adapted from \cite{DeLong1985}.}
	\label{Figure_adaptation}       
\end{figure}

Many organisms do not live at constant temperature or pressure, e.g., bacteria or cold-blooded animals. The membranes of their cells have the possibility to adapt to growth conditions.
Fig. \ref{Figure_adaptation} (left) shows the melting peaks of \textit{E. coli} membranes grown at two different temperatures, 15$^\circ$C and 37$^\circ$C, respectively \cite{Heimburg2007a}. One can recognize that the protein peaks recorded under different experimental conditions are found at the same temperatures indicating that protein structures are unaffected by growth temperature. However, the lipid peak (grey-shaded) adjusts to growth conditions. Since the lipids have no genome, this is most likely due to an adaption of the lipid composition to growth condition (see, e.g., \cite{Hazel1979, Hazel1995, Avery1995, vandeVossenberg1999}) and not due to selection of a particular strand. 

Membrane composition also adapts to growth pressure. Some bacteria live at very high pressures up to 1100 bars in the deep sea.  One example is the barophilic deep sea bacterium CNPT3 Fig. \ref{Figure_adaptation} (right). As a consequence of adaptation, the ratio of unsaturated to saturated lipids changes linear with growth pressure \cite{DeLong1985}. At higher pressures, one finds a higher fraction of unsaturated lipids, which display lower melting temperatures as saturated lipids. Since pressure increases transition temperatures due to the excess volume of the lipids in transitions, the adaptation of lipid composition in barophilic bacteria compensates for the pressure changes such that the function of the membrane is maintained.

It has further been shown that the lipid composition of \textit{E. coli} adapts to the presence of organic solvents (among those the anesthetic chloroform) in the growth medium \cite{Ingram1977}. This indicates that cells are `eager' to maintain a particular state of the membrane. The physiological temperature is practically always found at the upper end of the melting transition (an exception being the purple membrane \cite{Jackson1978}), and changes in the physiological parameters are compensated by adaptation of the lipid membrane composition. The functional purpose of the transitions will be discussed below.

\section{Influence of anesthetic drugs on membrane transitions}
\label{sec:Anesthesia}
\begin{figure}[b!]
	\includegraphics[width=11.5cm]{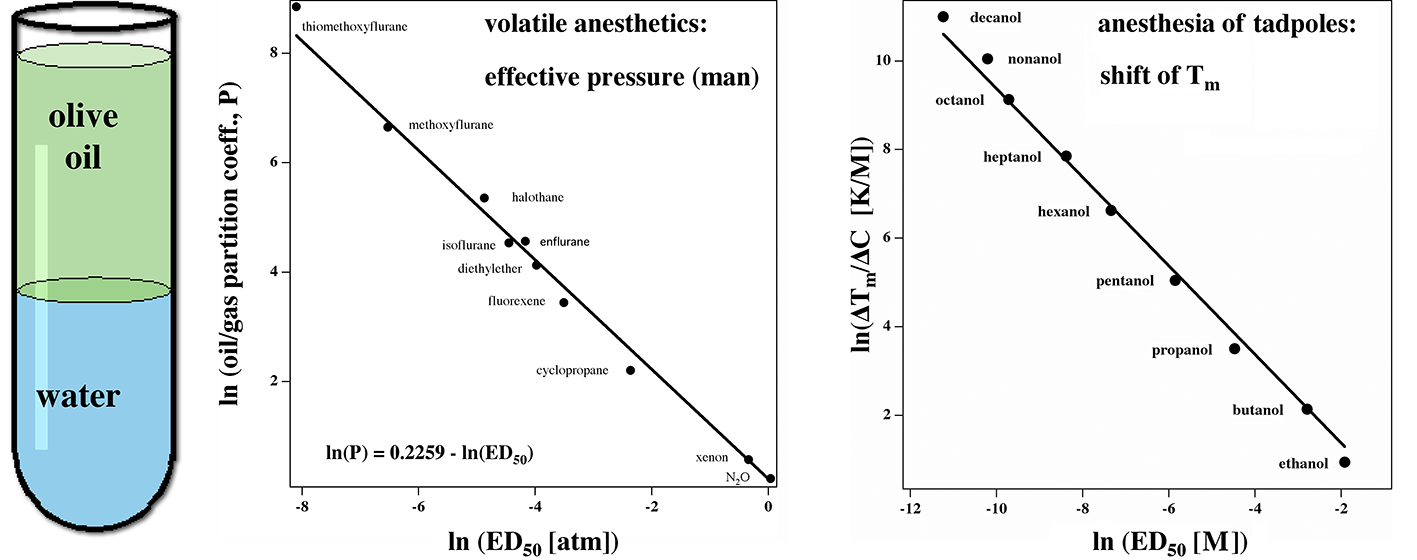}
	\caption{The Meyer-Overton correlation. Left: The partition coefficient of an anesthetic drug represents the ratio of the concentration of an anesthetic in equal amounts of oil and water. Center: When plotting the partition coefficient of general anesthetics against the critical dose of an anesthetic, ED$_{50}$, one obtains a straight line with a slope of -1 (in a double-logarithmic plot). Data adapted from \cite{Overton1991}. Right: The reduction of the membrane melting temperature as a function of the critical anesthetic dose for a series of alcohols. One also finds a slope of -1. Adapted from \cite{Kharakoz2001}}
	\label{Figure_Anesthetics}       
\end{figure}

General anesthetics represent a class of small drugs with a surprisingly different chemical structures, ranging from the noble gas Xenon (Xe), laughing gas (N$_2$O), diethylether (C$_2$H$_5$)$_2$O) and many alcohols to more complex molecules such as propofol (C$_{12}$H$_{18}$O) or thiopental (C$_{11}$H$_{18}$N$_2$O$_2$S). Surprisingly they all have something in common. Despite their very different chemical nature, they display generic properties summarized by the famous Meyer-Overton correlation \cite{Overton1991}. It states that the solubility of anesthetic drugs in membranes is inversely proportional to the critical dose that causes anesthesia. It is shown for numerous volatile anesthetics in Fig. \ref{Figure_Anesthetics} \cite{Overton1991}. The correlation can be written as \cite{Heimburg2007c}
\begin{equation}\label{Overton}
P\cdot [ED_{50}]=\mbox{const.}
\end{equation}

Here, $P$ is the oil/gas or the oil/water partition coefficient (i.e., the solubility of the drug in the lipid membrane), and [$ED_{50}$] is the effective dose of anesthetics where 50\% of all individuals are anesthetized. This correlation is remarkable but not well understood. Overton \cite{Overton1901} himself noted that the complete unspecificity of this correlation indicates that its origin is not based on molecular chemistry but rather on generic physical laws. An interpretation of the Meyer-Overton finding is complicated by the fact that there exist membrane-soluble molecules that do not serve as general anesthetics, e.g., cholesterol \cite{Cantor1997a, Cantor1997b} or long-chained alcohols \cite{Pringle1981, Kaminoh1992, Kamaya1984, Kamaya1986, Graesboll2014}. Thus, it is not exactly clear what distinguishes an anesthetic drug from other molecules. For this reason, the Meyer-Overton correlation has to be considered an empirical finding without much explanatory power. However, after a slight modification of the original assumption it obtains a profound physical meaning, and the requirement for an anesthetic drug become evident. 

There exists a generic physical-chemistry law called 'freezing-point depression' law. Some substances such as NaCl reduce the melting point of ice proportional to their molar concentration. This effect exists also for other salts, sugars or alcohols and is independent of chemical structure of the antifreeze agent. The lowering of the melting point of ice, $\Delta T_m$, is given by
\begin{equation}\label{freezingpoint}
\Delta T_m=-\frac{RT_m^2}{\Delta H}x_A\quad,
\end{equation}
where $T_m$ is the melting temperature of ice, $\Delta H$ is the heat of melting of ice, $R$ is the gas constant, and $x_A$ is the molar fraction of the molecule $A$ in water. When deriving this law, the main assumption is that the molecule $A$ is ideally soluble in water and but absolutely insoluble in ice. 
\begin{figure}[htb!]
	\includegraphics[width=11.5cm]{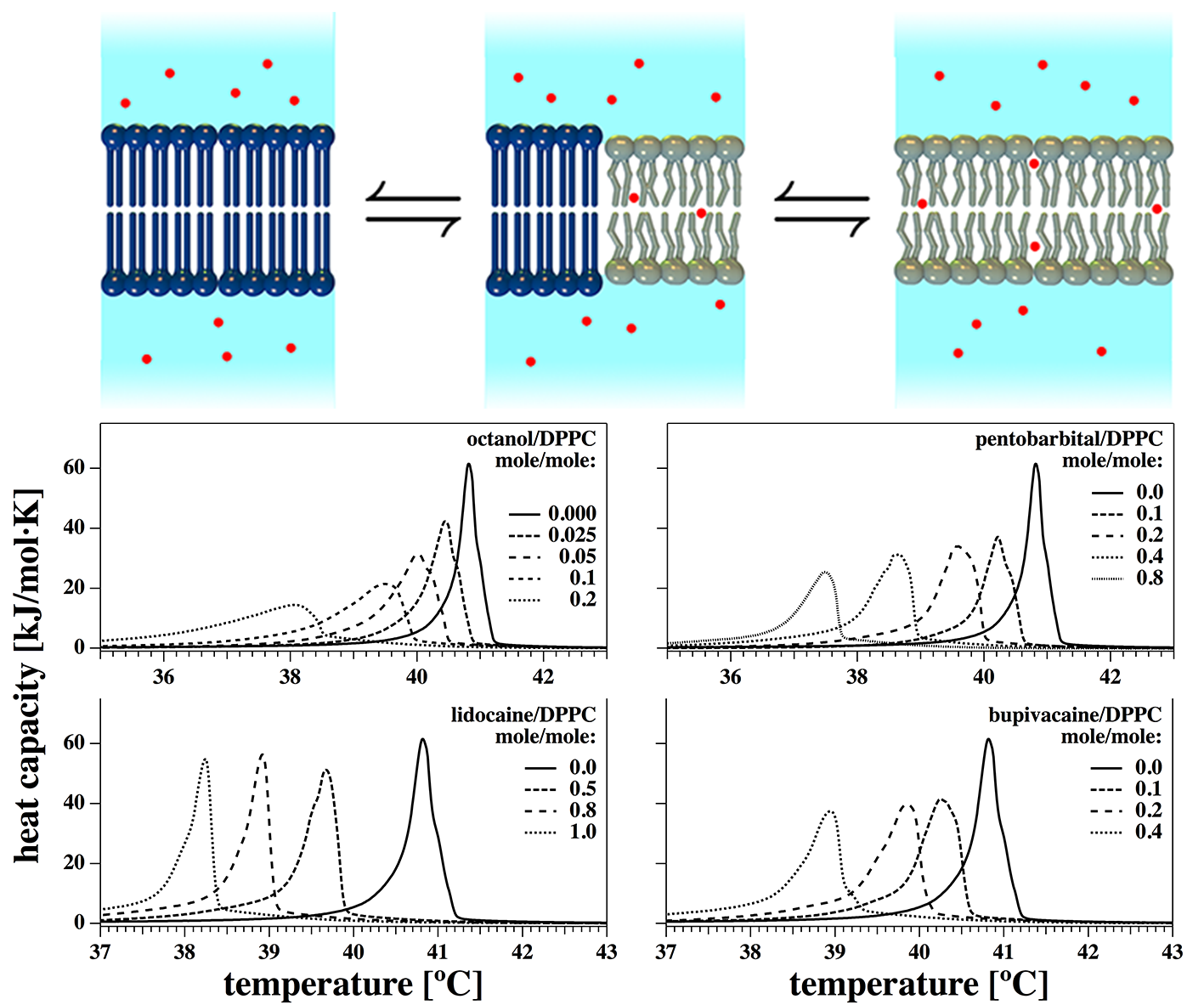}
	\caption{Anesthesia. Top: The melting-point depression law requires that anesthetic drugs (red dots) are ideally soluble in the liquid membrane but insoluble in the solid membrane. This is shown for three different temperatures (different liquid fractions). Bottom: Melting point depression for four different anesthetics at different concentrations. The alcohol octanol and the barbiturate pentobarbital are general anesthetics, and lidocaine and bupivacaine are local anesthetics. From \cite{Graesboll2014}.}
	\label{Figure_Anesthetics_B}       
\end{figure}

It has been shown that this correlation applies also to anesthetics and the melting points of lipid membranes \cite{Heimburg2007c, Graesboll2014}. In eq. (\ref{freezingpoint}), $\Delta H$ is the heat of membrane melting, and $x_A$ is the molar fraction of anesthetics dissolved in the membrane. General anesthetics lower the melting point of lipid membranes. One can calculate the lowering of the membrane melting temperature by postulating  that the drugs are soluble in the liquid membrane but not in the crystalline solid membrane - in exact analogy to the freezing-point depression law.  Interestingly, this finding is not only true for general anesthetics (e.g., octanol) but also for barbiturates (e.g., pentobarbital) and local anesthetics (such as lidocaine of bupivacaine) \cite{Graesboll2014}. This is shown in Fig.~\ref{Figure_Anesthetics_B}. Therefore, the Meyer-Overton correlation can be reformulated: "The effectiveness of an anesthetic drug is proportional to its liquid phase solubility, $P_f$, while it is insoluble in the solid membrane". Since $P_f\cdot [ED_{50}]$ corresponds the concentration of the anesthetic in the fluid membrane, which is constant at critical dose (eq. \ref{Overton}), one finds that the molar fraction of anesthetics in a membrane at critical dose is 2.6 mol\% \cite{Heimburg2007c}.  
Under these (and only under these) circumstances, the change of the melting temperature upon increase of drug concentration, $d \Delta T_m/d [C]$, is inversely proportional to the critical dose of anesthetics, $ED_{50}$, or
\begin{equation}\label{Overton2}
\frac{d \Delta T_m}{d [C]}\cdot [ED_{50}]=\mbox{const.}\;,
\end{equation}
From eq. \ref{Overton2} combined with the Meyer-Overton correlation for tadpole anesthesia, one can estimate the change in melting temperature at critical dose and finds $\Delta T_m=-0.6 ^\circ$ \cite{Heimburg2007c}, completely independent of the chemical nature of the general anesthetic. 

As mentioned above, the melting point depression is also valid for local anesthetics. This implies that local anesthetics have an identical influence on the thermodynamics of cell membranes as general anesthetics \cite{Graesboll2014}. Further, the melting-point depression law defines clear criteria for what is an anesthetic and what is not. Required is ideal solubility in the fluid membrane, and no solubility at all in the gel membrane. Only under these conditions the maximum melting-point depression can be achieved. Therefore, whenever the parameters of a particular drug fall on a linear line defined by  $d T_m/d[C]$ versus the partition coefficient $ED_{50}$ (Fig.\,\ref{Figure_Anesthetics}, right), the drug has a high potential of being a good anesthetic. These criteria are not fulfilled by molecules such as cholesterol and long-chain alcohols \cite{Kamaya1984, Kamaya1986, Graesboll2014}. Therefore, they are no good anesthetics. However, the criteria are fulfilled for all drugs that follow the Meyer-Overton correlation \cite{Heimburg2007c}, and for the barbiturates and local anesthetics investigated by us \cite{Graesboll2014} (Fig.\,\ref{Figure_Anesthetics_B}).

\section{Pressure reveral of anesthesia}
\label{sec:PressureReversal}

It has been shown by various authors that general anesthesia can be reversed by hydrostatic pressure \cite{Johnson1942, Johnson1950, Johnson1970, Halsey1975}. This surprising effect can be understood when considering that the melting temperature of lipid membranes depends on pressure, and that the melting-point depression of general and local anesthetics can be reversed by hydrostatic pressure \cite{Johnson1970, Trudell1975, Mountcastle1978, Kamaya1979, Galla1980}.
\begin{figure}[t]
	\sidecaption[htb!]
	\includegraphics[width=6cm]{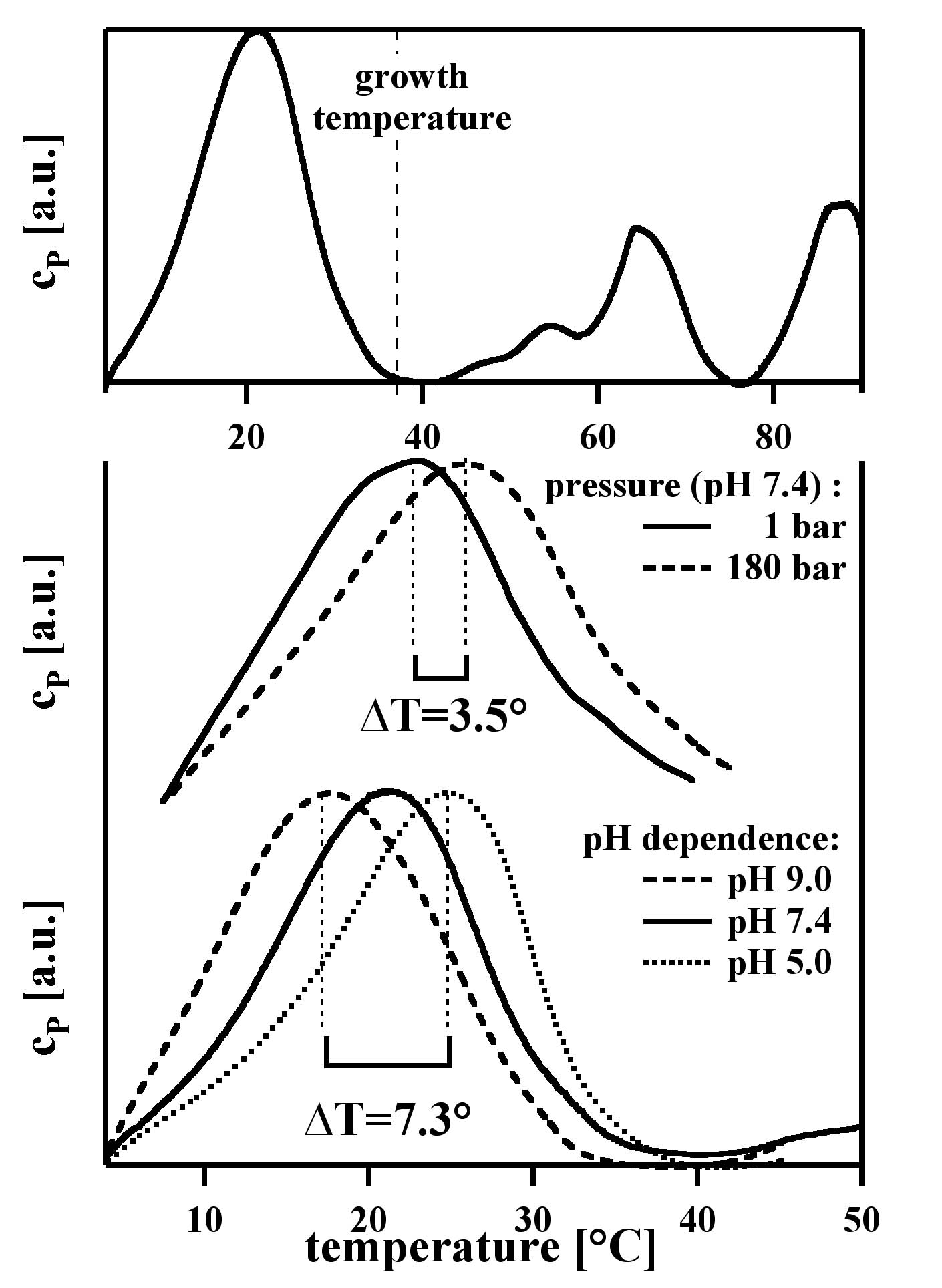}
	\caption{Top: Melting profile of \textit{E. coli} membranes. Center: Pressure dependence of the lipid meting peak of \textit{E. coli} membranes. Bottom: pH dependence of the lipid melting peak of of \textit{E. coli} membranes. Adapted from \cite{Heimburg2007c}.}
	\label{Figure_Ecoli_pressure_pH}       
\end{figure}
At the melting temperature of a lipid membrane, the molar free energy difference between liquid and solid membrane is zero ($\Delta \mu_0=\Delta H-T\cdot \Delta S = 0$). This leads to an equation for the melting temperature
\begin{equation}\label{pressure1}
T_m=\frac{\Delta H}{\Delta S}\;.
\end{equation}
The enthalpy is given by $\Delta H=\Delta E+p\Delta V$ and is therefore pressure dependent. This can be written as $\Delta H^{p}=\Delta H^{1 bar}\cdot (1+\gamma_V\Delta p)$, where $\Delta p$ is the pressure difference relative to 1 bar, and $\gamma_V=7.8\cdot 10^{-10}$ m$^2$/N is an empirical parameter measured for various artificial lipids and lung surfactant \cite{Ebel2001}. The change in melting temperature at pressure $p$ is now given by $\Delta T_m^p$:
\begin{equation}\label{pressure2}
\Delta T_m^p=\gamma_V\cdot \Delta p\cdot T_m^{1 bar}\;.
\end{equation}
This implies that it increases linearly with pressure. The influence of pressure on lipid melting transitions has been studied in detail for artificial lipids in \cite{Winter1989, Ebel2001}. However, the effect also exists in biological membranes, such as lung surfactant \cite{Ebel2001} or \textit{E. coli}-membranes \cite{Heimburg2007c}, shown in Fig. \ref{Figure_Ecoli_pressure_pH} (center). The effect of pressure on biological preparations and on artificial membranes displays a very similar order of magnitude.
Using eqs. \ref{freezingpoint} and \ref{pressure2}, one can calculate the pressure necessary to reverse general anesthesia. One obtains
\begin{equation}\label{pressure3}
\Delta p=\frac{R T_m^{1 bar}}{\gamma_V\Delta H^{1 bar}}x_A \;.
\end{equation}

At this pressure, the melting point depression is has exactly the same magnitude but opposite sign as the pressure induced shift.
For the parameters of the artificial lipid dipalmitoyl phosphatidycholine ($T_m=314$ K and $\Delta H\approx 35 kJ/mol$ and a critical fraction of 2.6 mol\%), this happens for approximately 25 bars. Values in the literature are of the same order but vary a lot. Johnson and Flagler \cite{Johnson1950} found that tadpoles anesthesia in 3-6 \% of ethanol (3-6 times of critical dose$ED_{50}$) could be reversed with 140-350 bars of pressure. Our calculation above would lead to a pressure of 75-150 bars of pressure. This is remarkably close taking into account that this calculation is based on the parameters of an artificial lipid while the pressure reversal was measured on tadpoles.

\section{Influencing the melting transition by pH, sodium, potassium, calcium and other charged ligands}
\label{sec:Influence}
\begin{figure}[htb]
	\sidecaption[t]
	\includegraphics[width=7cm]{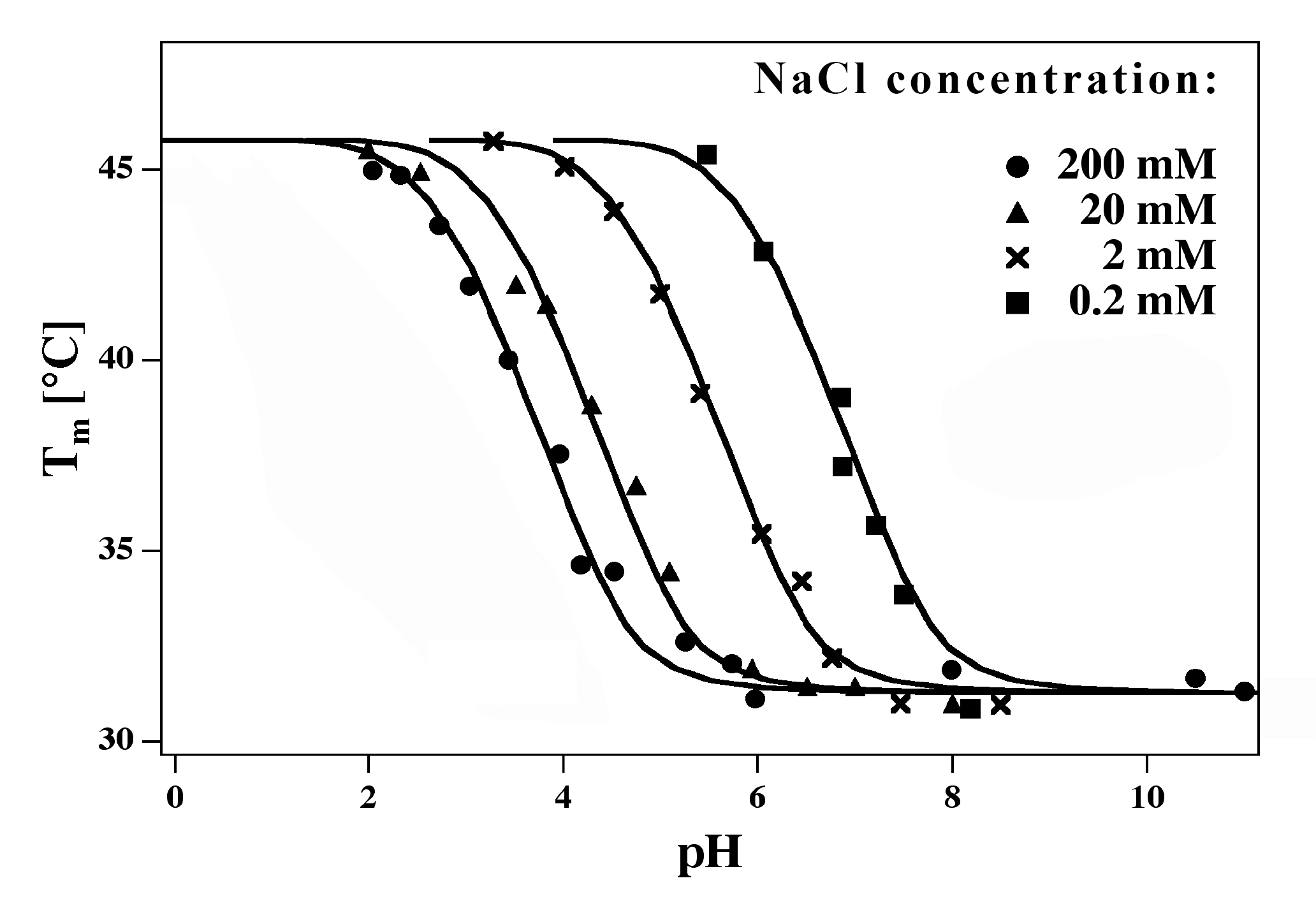}
	\caption{The melting temperature of methylphosphatidic acid (MPA) as a function of pH and NaCl concentration. Adapted from \cite{Trauble1976, Heimburg2010}.}
	\label{Figure_Tm_pH_NaCl}       
\end{figure}

\paragraph{\textbf{Electrostatics and the influence of ions}}
Biological membranes contain negatively charged lipids, which are usually asymmetrically distributed over the inner and outer monolayer of the membrane \cite{Rothman1977, Rothman1977b}. This implies that the membranes possess both an electrical potential, and a net polarization across the membrane. The negative charges lead to repulsive forces in the membrane plane, that cause a lowering of the melting temperature of the membrane \cite{Heimburg1996b}. The electrostatic potential can be partially shielded by protons (H$^+$), Na$^+$- or K$^+$-ions and Ca$^{2+}$-ions. Therefore, an increase in NaCl (or CaCl$_2$) concentration, or a lowering of the pH leads to an increase of the melting temperature \cite{Trauble1976} (see Fig. \ref{Figure_Tm_pH_NaCl}). An influence of the melting transition can also be induced by the binding of basic proteins such as cytochrome c \cite{Heimburg1996b}.

\paragraph{\textbf{Influence of pH on anesthesia}}
This influence of ions and large molecules on the electrostatics of membranes, and the resulting changes in transition temperature of membranes is likely to have an influence on anesthesia as described in sections \ref{sec:Anesthesia} and \ref{sec:PressureReversal}. Since lowering of pH increases and anesthetic drugs decrease the transition temperature in membranes, it is likely the decreasing pH reverses the anesthetic effect. In fact, it is known that inflammation leads to the failure of anesthesia. The pH of inflamed tissue is lowered by about 0.5 units \cite{Punnia-Moorthy1987} which is believed to be the cause for the failure. It has been shown in \cite{Heimburg2007c} that the increase of the transition caused by a pH-drop of this order is sufficient to reverse the melting-point depression caused by anesthetics (Fig.\,\ref{Figure_Ecoli_pressure_pH}, bottom).

\section{Elastic Constants and time scales}
\label{sec:Elastic}
\paragraph{\textbf{Elastic constants}}
In the transition of a membrane, its enthalpy, volume and area are changing. The temperature dependence of both volume and area is given by \cite{Heimburg1998, Ebel2001, Pedersen2010}
\begin{equation}\label{eq:elastic_1}
\Delta V(T)=\gamma_V\cdot\Delta H(T) \qquad\mbox{and}\qquad  \Delta A(T)=\gamma_A\cdot\Delta H(T)
\end{equation}
This is an empirical finding from experiment. Simultaneously, it is known from the fluctuation dissipation theorem that the heat capacity is proportional to the mean square fluctuations in enthalpy \cite{Heimburg1998},
\begin{equation}\label{eq:elastic_2}
\Delta c_p=\frac{\left\langle \Delta H^2\right\rangle -\left\langle \Delta H\right\rangle^2}{R T^2} \;,
\end{equation}
while the isothermal volume and area compressibilities are proportional to the mean square fluctuations in volume and area \cite{Heimburg1998}, respectively 
\begin{equation}\label{eq:elastic_3}
\Delta \kappa_T^V=\frac{\left\langle \Delta V^2\right\rangle -\left\langle \Delta V\right\rangle^2}{\left\langle V\right\rangle R\cdot T} \qquad\mbox{and}\qquad\Delta \kappa_T^A=\frac{\left\langle \Delta A^2\right\rangle -\left\langle \Delta A\right\rangle^2}{\left\langle A\right\rangle R\cdot T}\;.
\end{equation}
\begin{figure}[bt]
	\sidecaption[t]
	\includegraphics[width=7cm]{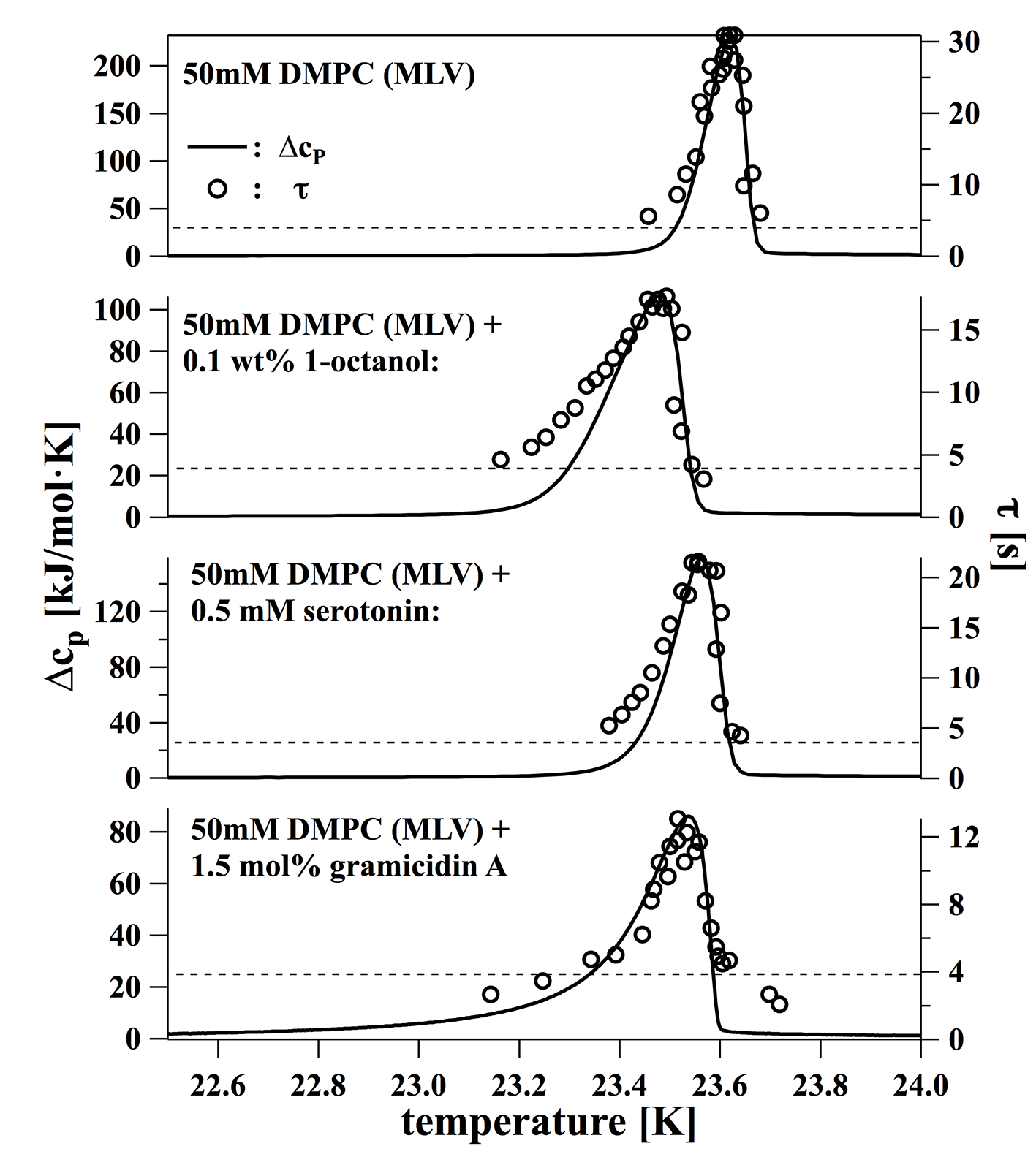}
	\caption{Relaxation time scale of membranes in the absence and presence of the general anesthetic octanol, the neurotransmitter serotonine, and the antibiotic peptide gramicidin A, as compared to the heat capacity (solid lines). The relaxation times are at maximum in the transition. Adapted from \cite{Seeger2007}.}
	\label{Figure_Relaxation}       
\end{figure}

From eqs.(\ref{eq:elastic_1})-(\ref{eq:elastic_3}) follows that
\begin{equation}\label{eq:elastic_4}
\Delta \kappa_T^V=\frac{\gamma_V^2\cdot T}{\left\langle V\right\rangle }\Delta c_p \qquad\mbox{and}\qquad \Delta \kappa_T^A=\frac{\gamma_A^2\cdot T}{\left\langle A\right\rangle }\Delta c_p \;,
\end{equation}
i.e., volume and area compressibility are proportional to the heat capacity, and can be determined from calorimetric data. A similar relationship can be derived for the bending elasticity \cite{Heimburg1998}:
\begin{equation}\label{eq:elastic_5}
\kappa_{bend}=\frac{16\gamma_A^2\cdot T}{\left\langle D\right\rangle ^2 \left\langle A \right\rangle }\Delta c_p
\end{equation}
This implies that membranes are much softer and more flexible close to a transition.

\paragraph{\textbf{Relaxation times}}
Further, by a slightly more sophisticated argument it can be shown that the relaxation time $\tau$ is proportional to the heat capacity \cite{Grabitz2002, Seeger2007}
\begin{equation}\label{eq:elastic_6}
\tau=\frac{T^2}{L}\Delta c_p
\end{equation}
where $L$ is a constant. This relation describes the time it takes for a membrane to equilibrate after a perturbation. Due to arguments originating from Lars Onsager \cite{Onsager1931b}, it is also the lifetime of the fluctuations. We show in Fig. \ref{Figure_Relaxation} that the relaxation time is roughly proportional to the heat capacity. At the melting transition, fluctuations are especially slow, an effect known as critical slowing-down. Addition of drugs such as octanol (a general anesthetic), serotonine (a neurotransmitter) and gramicidin A (an antibiotic peptide) all influence the heat capacity profile and thereby influence relaxation times by a purely physical mechanism (Fig. \ref{Figure_Relaxation}). We will show the effect of this below in the section treating the lifetime of lipid pores (lipid channels) in membranes. 

Summarizing, both volume and area compressibility, bending elasticity and relaxation time are proportional to the heat capacity and can therefore be estimated from the calorimetric experiment. All of these functions display maxima in the transition. Thus, it is clear that the melting transitions have a profound influence on the mechanical properties of membranes, and on the equilibration timescales.

\section{Voltage, capacitance and transitions}
\label{sec:Capacitance}
\begin{figure}[b]
	\includegraphics[width=11.5cm]{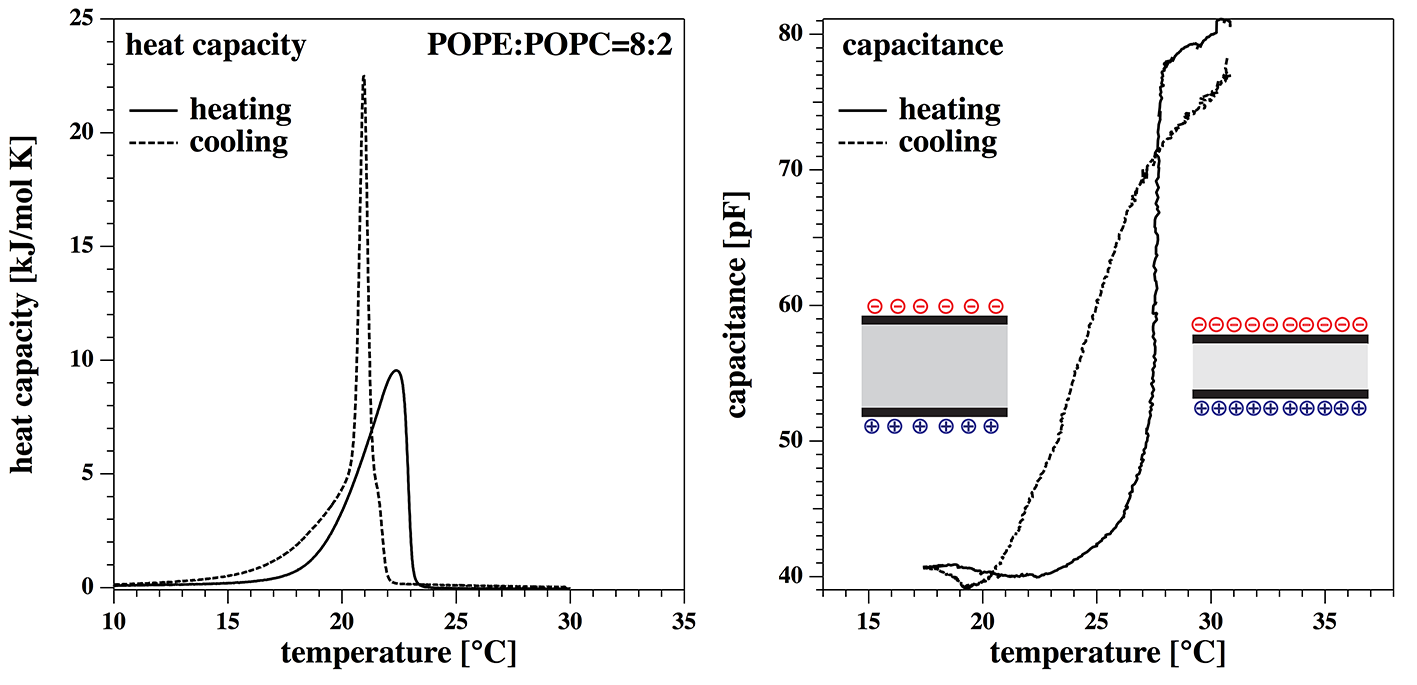}
	\caption{Change in capacitance of an artificial DOPE:DOPC=8:2 membrane patch. From \cite{Zecchi2017}.}
	\label{Figure_Capacitance}       
\end{figure}

As mentioned in the previous section, lipid membranes change their dimension in the melting transition. For instance, the lipid DPPC decreases its thickness $D$ from 4.79 nm to  3.92 n, and increases its area $A$ from 0.474 nm$^2$ (at 25$^\circ$) to 0.629 nm$^nm$ (at 50$^\circ$) \cite{Heimburg1998}. At the transition itself, this corresponds to a reduction in relative thickness of -16.3\% and and increase in relative area of 24.6\%. This has a profound influence on the membrane capacitance \cite{Heimburg2012}. Its value is given by 
\begin{equation}\label{eq:capacitance}
C_m=\varepsilon_0 \varepsilon \frac{A}{D}
\end{equation}
where $\varepsilon_0=8.854\cdot10^{−12}$ F/m is vacuum permittivity, $\varepsilon\approx 3$ and is the dielectric constant of the membrane core. With the above constants, the capacitance is about $0.6 \mu F/cm^2$ at 25$^\circ$, very close to reported values in the biological literature. However, the membrane capacitance of DPPC increases by about 50\% in the transition. Fig. \ref{Figure_Capacitance} shows a measurement of the capacitance of a POPE:POPC=8:2 lipid mixture as a function of temperature \cite{Zecchi2017}. The capacitance changes by roughly 100\% upon going through the transition. While this effect is normally neglected, it represents a significant change that has to be taken into account in modeling of the electrical properties of biomembranes, e.g., of nerves.

\begin{figure}[t]
	\sidecaption[t]
	\includegraphics[width=7cm]{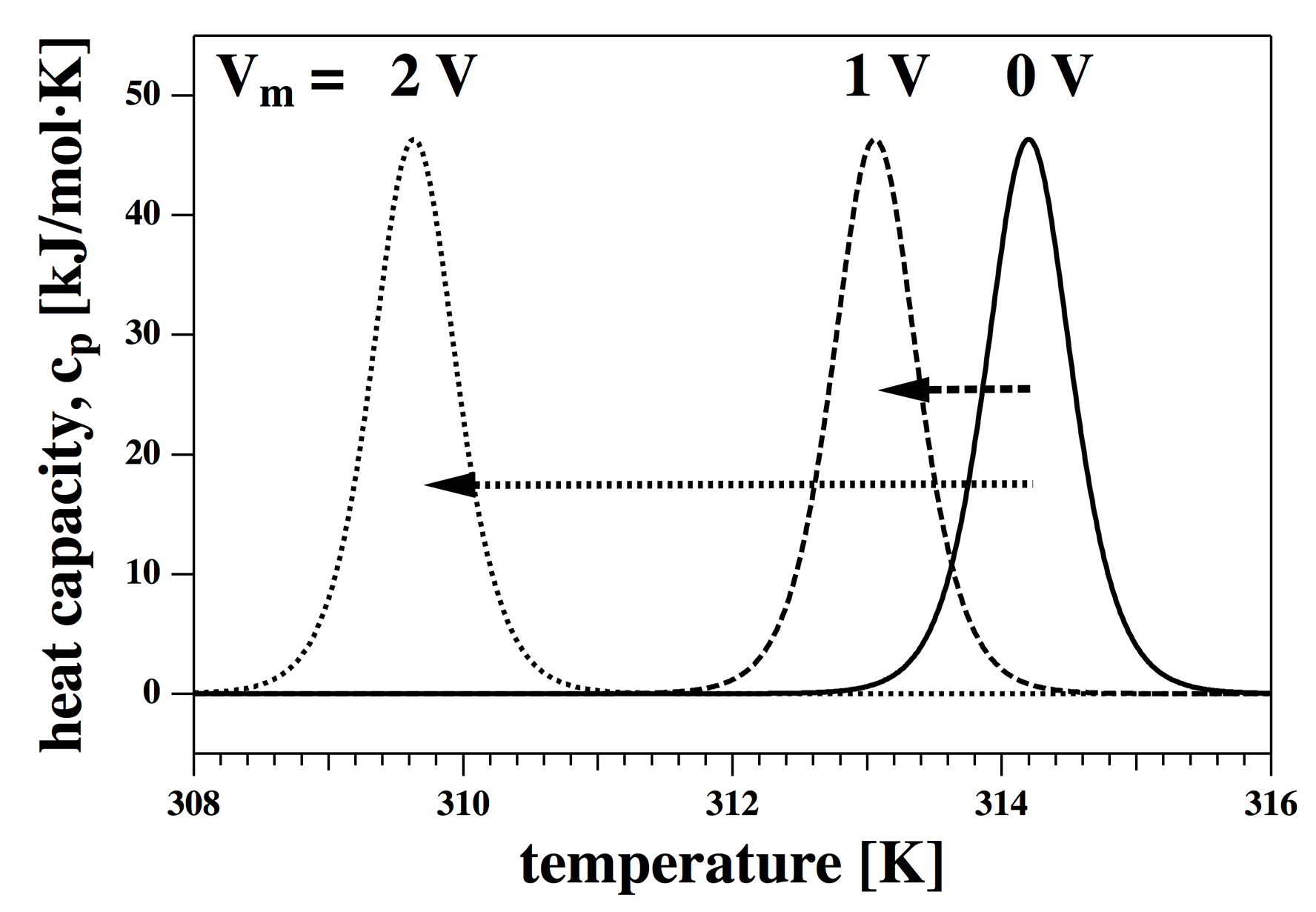}
	\caption{Calculated change of the transition temperature of DPPC on voltage for $V_0=0$\,V. From \cite{Heimburg2012}.}
	\label{Figure_Voltage_B}       
\end{figure}
The fact that different lipid phases display different capacitance renders the membrane transition voltage dependent. It has been shown in \cite{Heimburg2012} that the transition temperature displays a quadratic dependence on voltage on $V-V_0$, where $V_0$ is the equilibrium polarization of the membrane:
\begin{equation}\label{eq:voltage_Tm}
T_m(V)=T_{m,0}\left(1+\alpha \left(V-V_0\right)^2\right) \;,
\end{equation}
where the coefficient $\alpha=-0.003634$ 1/V$^2$ was calculated for the lipid DPPC. For the case of $V_0=0$ V (no equilibrium polarization of the membrane), the dependence of the transition on the voltage is shown in Fig. \ref{Figure_Voltage_B}. If $V_0 \ne 0$ V, the voltage dependence for positive and negative voltages is different.

\section{Permeability and Channels}
\label{sec:Permeability}

\begin{figure}[b]
	\includegraphics[width=11.5cm]{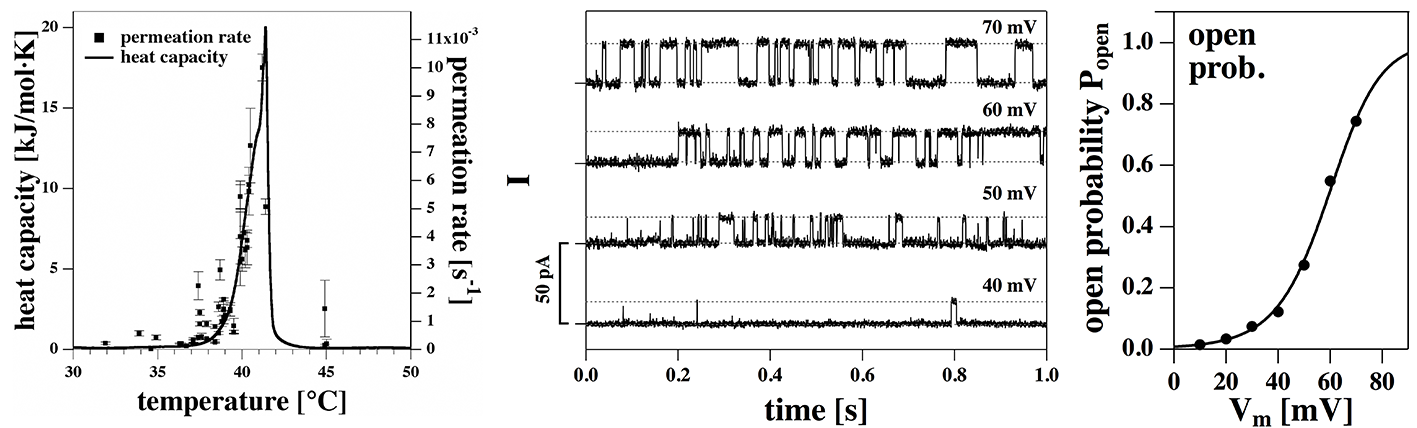}
	\caption{Left: The permeability of an artificial lipid membrane for a fluorescence marker. From \cite{Blicher2009}. Center: The channels in a patch recording of synthetic lipid membrane are voltage gated. Right: Open-probability of lipid channels as a function of voltage. From \cite{Blicher2013}.}
	\label{Figure_pore_opening}       
\end{figure}
Close to transitions, artificial membranes are no electrical insulators but display a maximum in membrane permeability \cite{Papahadjopoulos1973, Sabra1996, Blicher2009} (Fig. \ref{Figure_pore_opening}, left). This implies that close to transitions an important prerequisite of many membrane theories based on ion channel proteins (i.e., that ions do not pass through the lipid membrane) does not hold true. Further, it has been shown that close to transitions, artificial lipid membranes can display channel opening-events very similar to those reported for channel proteins (Fig. \ref{Figure_pore_opening}, center) \cite{Blicher2009, Laub2012}. It is believed that the channels are related to transient opening of hydrophilic or hydrophobic pores in the membrane (for a review see \cite{ Glaser1988, Boeckmann2008, Heimburg2010, Mosgaard2013b}). These lipid channels have been shown to display very similar single-channel conductances as proteins, and quite comparable open-lifetime distributions \cite{Gallaher2010, Blicher2013}. They can be voltage-gated (\ref{Figure_pore_opening},center) and current-voltage relations can display outward rectification \cite{Blicher2013, Mosgaard2013b, Mosgaard2015a}. This is particularly interesting because it is generally believed that both features are a property of proteins. It is clear from membrane experiments that this assumption is not correct.

\begin{figure}[t]
	\sidecaption[t]
	\includegraphics[width=11.5cm]{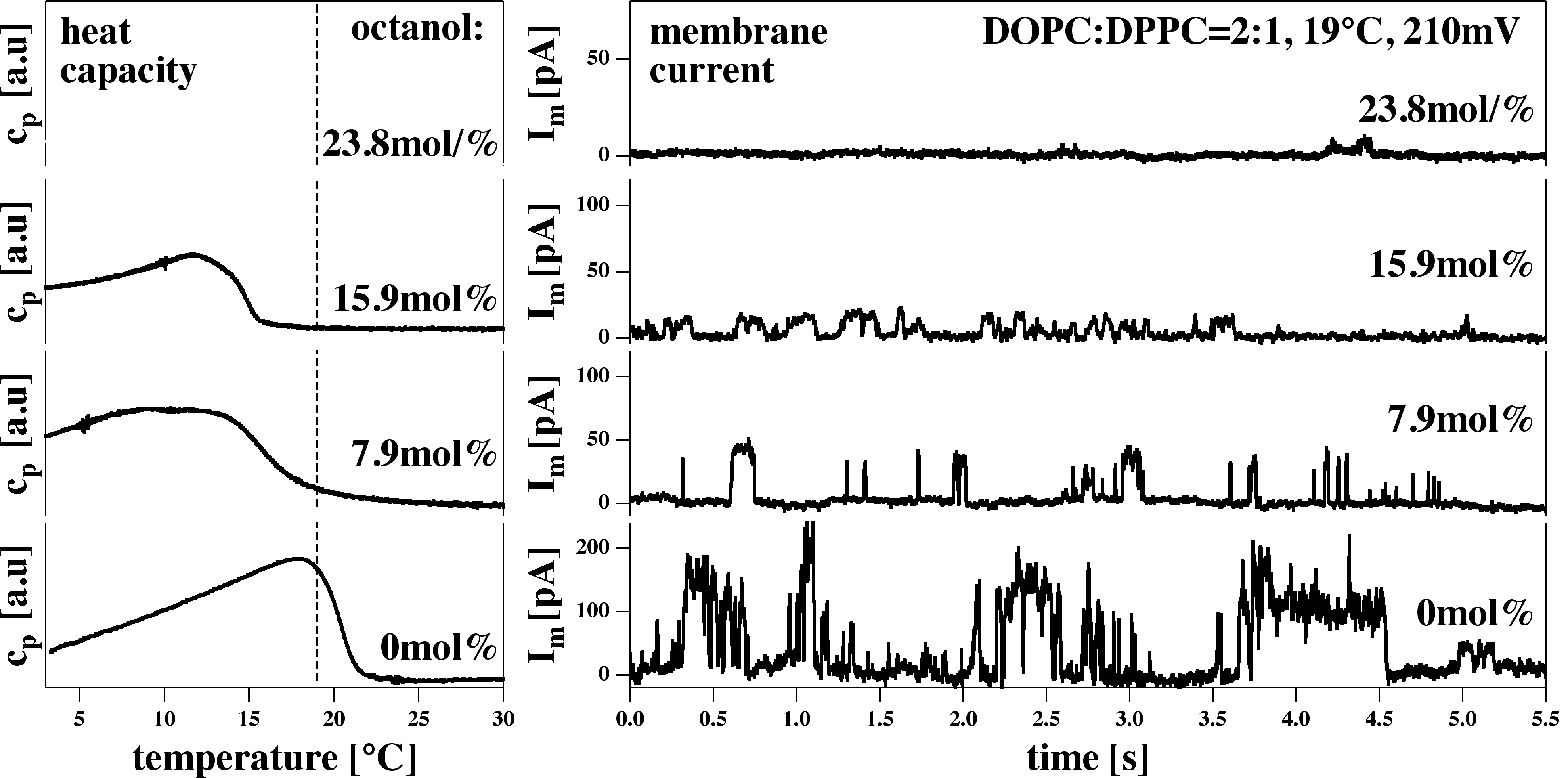}
	\caption{The influence of anesthetics on the channel activity of a pure lipid membrane close to its transition.The vertical dashed lines indicate experimental temperature. Left: Heat capacity profiles. Right: Channel traces for different amounts of anesthetics. The channel conductance changes and the open probability decreases until the lipid channels are completely "blocked". From \cite{Blicher2009}.}
	\label{Figure_Lipid_channel_anesthetics}       
\end{figure}
Since the permeability is at a maximum close to a transition, it is not surprising that membrane-active drugs can influence the permeability of the biomembranes \cite{Sabra1996, Blicher2009}. Fig. \ref{Figure_Lipid_channel_anesthetics} displays a membrane measured at the transition temperature of a lipid mixture (dashed lines in the $c_p$ profiles). Consecutive addition of a general anesthetics (octanol) lowers the melting temperatures due to melting-point depression and moves the experimental situation towards lower temperatures - away from the maximum heat capacity situation. The right hand panels of Fig. \ref{Figure_Lipid_channel_anesthetics} show the conduction events from electrophysiological patch recordings. Both, single-channel conductance, open-probabilities of lipid pores as well as the open lifetime of the lipid channels are reduced upon addition of the anesthetic octanol. It is worth-wile noting that the pure lipid membrane reacts in a very similar manner to the general anesthetic octanol as some protein channels, e.g., voltage-gated sodium channels \cite{Horishita2008} and the acetylcholine receptor \cite{Zuo2001}. The channels are apparently `blocked'. However, in the case of a pure lipid membrane this cannot hold as an explanation since there are no molecular channels that can be blocked. Obviously, there exists a purely physical mechanisms that can prevent conduction events, which is related to the shift of transitions. Summarizing, one has to be cautious when channel-events recorded in electrophysiological experiments are interpreted: They could originate from fluctuating pores in lipid membranes, and the "blocking" of such channels by drugs can be due to the thermodynamics of the molecules dissolving in the membranes \cite{Blicher2009, Heimburg2010}. 

The concepts presented here also seem to apply to protein channels reconstituted in lipid membranes. Seeger et al. \cite{Seeger2010} demonstrated that the KcsA potassium channel displays a maximum conductance when the host lipid has a transition, and that the open lifetime of the channel is also at maximum. This indicates that the properties of the protein channel are closely tied to the thermodynamics of the surrounding lipid membrane (see also \cite{Mosgaard2013b}).

\section{Pulses along Membranes}
\label{sec:Pulses}
In the model for the nervous impulse established by A. L. Hodgkin and A. F. Huxley \cite{Hodgkin1952b, Johnston1995}, the axon is considered a cylindrical membrane. Cable theory is modified by employing a voltage-dependent membrane conductance for potassium and sodium. This effectively leads to the following differential equation for the time and position dependence of the voltage in an action potential:
\begin{equation}\label{eq:HH1}
	\frac{a}{2 R_i}\frac{\partial^2 V}{\partial x^2}=C_m\frac{\partial V}{\partial t}+g_K(V,t)\cdot(V-E_K)+g_{Na}(V,t)(V-E_{Na})+ \mbox{leak currents} \;.
\end{equation}
Here, all complications from the voltage-gating of ion channel proteins are hidden in the complicated time and voltage-dependent conductances for potassium and sodium, $g_K$ and $g_{Na}$, respectively. These were parametrized by \cite{Hodgkin1952b} from experiments on squid axons \cite{Hodgkin1952a} using many fit constants. The capacitance $C_m$ is assumed being constant. $E_K$ and $E_{Na}$ are the Nernst potentials that reflect the potassium and sodium concentrations inside and outside of the neuron. $R_i$ is the internal specific resistance of the cytoplasm, and $a$ is the cable radius. Eq. (\ref{eq:HH1}) displays pulse-like solutions, which are generally considered to be the explanation for the action potential.
\begin{figure}[htb]
	\includegraphics[width=11.5cm]{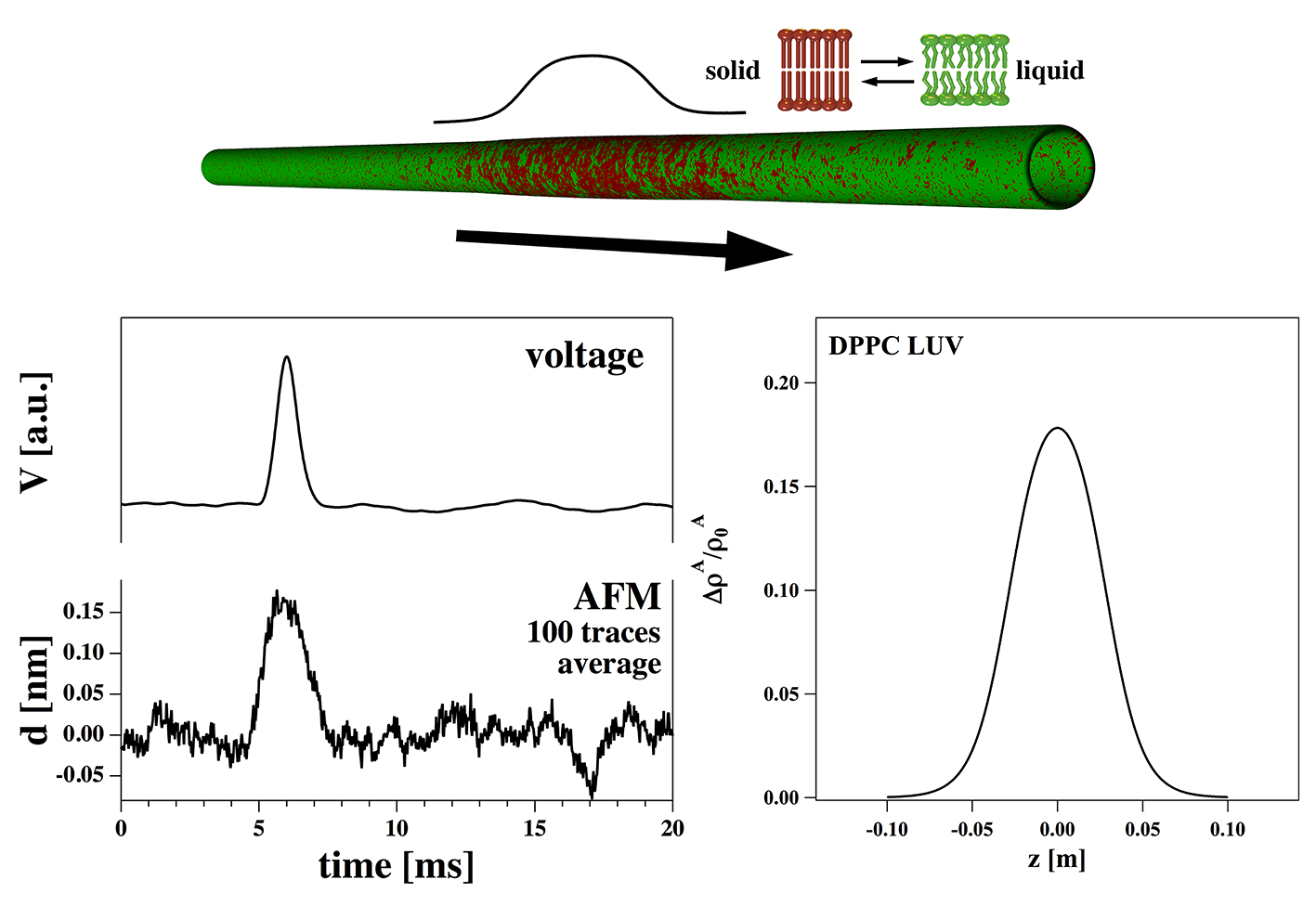}
	\caption{The nerve pulse as a electromechanical pulse involving phase transitions.Top: Schematic drawing of a propagating solitary pulse that involves a transition (solid=red, liquid=green) From \cite{GonzalezPerez2016}. Bottom left: Mechanical pulse (changes in membrane thickness) measured by atomic force microscopy as compared to the voltage pulse. Adapted from \cite{GonzalezPerez2016}. Right: Calculated pulse using the elastic parameters of membranes of the lipid DPPC. Adapted from \cite{Heimburg2005c}.}
	\label{Figure_solitons_1}       
\end{figure}

However, the presence of a phase transition in neural membranes gives rise to a very interesting related phenomenon: The possibility of propagating solitary pulses in cylindrical membranes. These pulses are electromechanical in nature and resemble the propagation of sound. They consist of a variation in membrane area density, $\Delta \rho$. They are the consequence of the change of the elastic constants close to transitions, and the frequency dependence of the sound velocity (known as dispersion). The differential equation for the membrane density changes can be written as (technical details can be found in \cite{Heimburg2005c, Lautrup2011})
\begin{equation}\label{eq:HH2}
\frac{\partial^2 \Delta \rho}{\partial t^2}=\frac{\partial}{\partial x}\left((c_0^2+p\Delta \rho + q\Delta \rho^2)\frac{\partial \Delta \rho}{\partial x}\right)-h\cdot \frac{\partial^4 \Delta \rho}{\partial x^4}
\end{equation}
Here, $c_0$ (speed of sound), $p$ and $q$ represent the elastic constants of the membrane close to a transition. They have been determined for artificial and biological membranes using the concepts leading to eq. (\ref{eq:elastic_4}) \cite{Heimburg2005c}. The dispersion parameter $h$ is responsible for describing the frequency dependence of the speed of sound. The theory behind eq. (\ref{eq:HH2}) is often called the "soliton"-theory. While this differential equation has a somewhat different structure, it also possesses pulse-like solutions. During the pulse, the membrane is transiently shifted from a liquid to a solid state and back. This is schematically demonstrated in Fig. \ref{Figure_solitons_1} (top) where it is shown that a solid region (red) travels  in a liquid environment (green). This goes along with changes in membranes thickness and capacitance. We have measured thickness changes by atomic force microscopy. Fig. \ref{Figure_solitons_1} (bottom left) shows the thickness changes in a lobster axon and the corresponding voltage changes. Both mechanical and voltage changes are exactly in phase \cite{GonzalezPerez2016}. In Fig. \ref{Figure_solitons_1} (bottom, right) a solution of eq. (\ref{eq:HH2}) is given. The density pulses compare well with the shape of both the voltage signal and the thickness change of the membrane. 
Similar findings have been made by \cite{Iwasa1980a, Iwasa1980b, Tasaki1989, Kim2007} on squid axons, garfish nerves and mammalian nerve terminals. There exists no reasonable doubt that the action potential possesses a mechanical component. Therefore, the action potential couples to the thermodynamics of the membrane. Eq. (\ref{eq:HH2}) describes many non-electrical features of nerves that are not contained in the Hodgkin-Huxley theory, including the mechanical changes, the length contraction, the reversible heat absorption observes in experiments (e.g., \cite{Ritchie1985}) and the observation that two colliding nerve pulses may pass through each other \cite{GonzalezPerez2014, GonzalezPerez2016}. 

\section{Influence of the thermodynamic variables on excitability.}
\begin{figure}[hbt]
	\includegraphics[width=11.5cm]{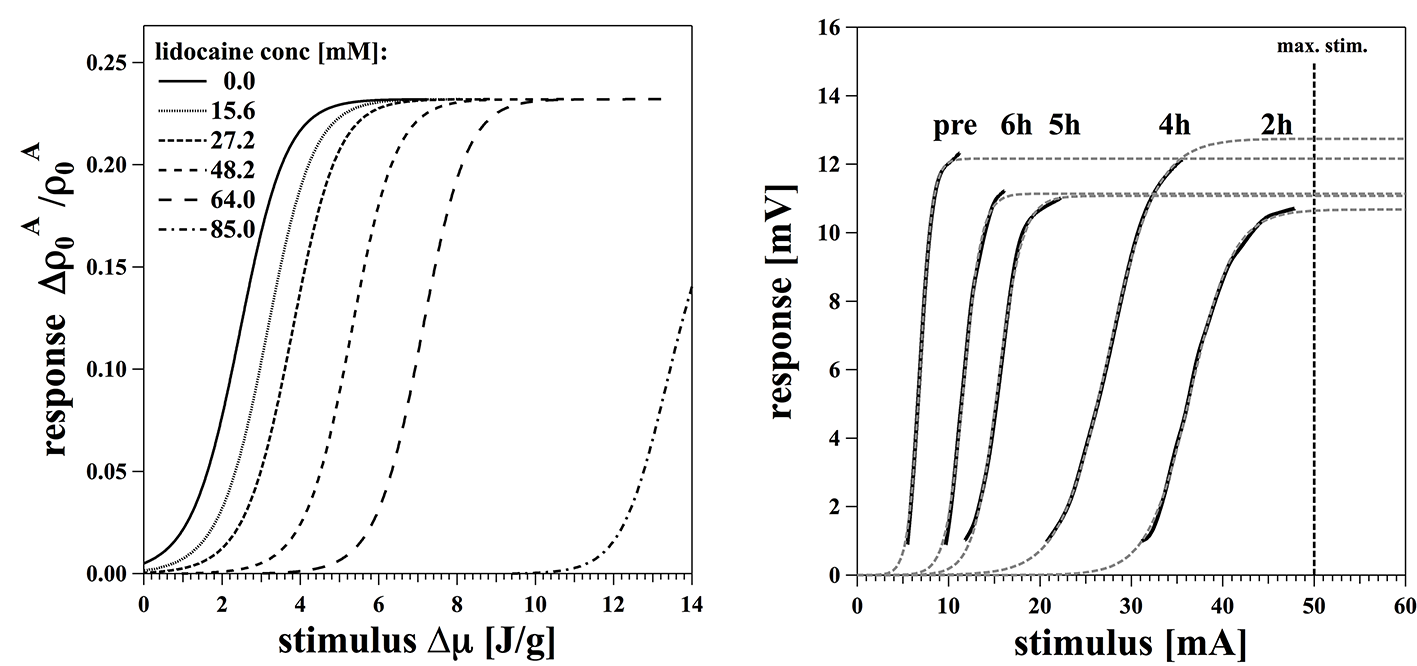}
	\caption{Stimulus-response curves. Left: Calculated density changes as a function of stimulation free energy for various lidocaine concentrations. Increasing amount of the anesthetic increases the stimulation threshold. The maximum amplitude is conserved, and no blocking of the nerve occurs. Right: Stimulus-response profiles for the human median nerve at various times after administration of lidocain. Adapted from \cite{Moldovan2014, Wang2018}.}
	\label{Figure_Stimulus_Response}       
\end{figure}

The beauty of a thermodynamic model for the nerve pulse related to membrane transitions lies in the possibility to influence the nerve pulse by changes in the thermodynamic variables. E.g., anesthetics move the melting transition further away from physiological temperature, and thereby make it more difficult to push the membrane through is phase transition. This implies that it becomes more difficult to stimulate a solitary pulse as described in the previous section. Nevertheless, it is clear that while in the soliton model anesthetics will increase the stimulation threshold, they will not block a nerve pulse. This is conceptually shown in Fig. \ref{Figure_Stimulus_Response} (left). Here, the free energy difference between liquid and solid membrane is plotted versus the membrane density for different concentrations of anesthetics (remember that the amplitude of the soliton represents a change in membrane density). One finds sigmoidal profiles. This plot describes the free energy necessary to stimulate a solitary pulse and how it is influenced by anesthetics. We define the stimulation threshold as the half-maximum density. It corresponds to the position of the melting transition. One can recognize a significant shift in threshold for increasing concentration of the anesthetic drug. 
These theoretical predictions are compared to experiments on the human median nerve (Fig. \ref{Figure_Stimulus_Response} (right), adapted from Moldovan et al. \cite{Moldovan2014}). In this experiment, the median nerve was locally anesthetized by high doses of lidocaine. The stimulus was administered be a current between two electrodes at the lower arm, and the response was measured in the thumb. The stimulus-response curve was recorded at different time after the experiment while the anesthetic concentration was gradually washed out. The stimulus-response profile continuously decreases its threshold, and returned to the original situation after about 24 hours. After this time, the stimulus-response curve was the same as before the experiment. The threshold increases by a factor of 7-10 without any obvious blocking of the nerve. The maximum amplitude stays about the same, just as in the calculation in the left hand panel. Assuming that the action potential can be explained by the progagation of a soliton in the membrane, it becomes immediately clear why anesthesia can be reversed by pressure: Pressure reduces the threshold by moving the phase transition temperature closer to physiological temperature. Thereby, it counteracts the effect of anesthetics on the stimulation threshold.

\section{Summary}
\label{sec:Summary}
In the past, the power of a thermodynamic treatment of biomembranes did not receive the appreciation it deserves. It is likely that many membrane-related properties are intrinsically related to the profound thermodynamic couplings described here. Treating phenomena such as propagating pulses, absorption of drugs, emergence of channels as independent processes will not elucidate the underlying mechanisms. In fact, separating these phenomena may even violate the second law of thermodynamics as shown in \cite{Heimburg2008}. However, the beauty of the thermodynamic couplings begins to attract more attention. In respect to biomembrane phenomena, it finds its application in the emergence of membrane pores as a consequence of thermal collisions, in propagating pulses as density variations close to regimes with large fluctuations, and the control of such processes by the variation of intensive variables including anesthetic drugs.

Here, we have reviewed the experimental evidences for transitions in biomembranes and the effect that transitions may have on heat capacity, elastic constants, relaxation lifetimes and on capacitance. A natural consequence of the transitions is the emergence of long-lived pores in the membrane. This effect makes the membrane most permeable in the transition. Various factors influence the transitions, e.g., pressure and voltage, but also the presence of drugs such as anesthetics. The latter lower transitions, while pressure increases them. 
Response functions (or susceptibilities) are derivatives of extensive thermodynamic variables with respect to intensive variables. These are often the properties accessible in an experiment. Examples are heat capacity ($dH/dT$), capacitance ($dq/d \Psi$), volume and area compressibility ($-V^{-1}dV/dp$ or $-A^{-1}dA/d\Pi$, respectively), the volume expansion coefficient ($dV/dT$) or the bending elasticity. All of them display maxima in transitions and are related to fluctuation relations.  Changes in the intensive variables (temperature, voltage, hydrostatic and lateral pressure, and the chemical potential of membrane components) influence the position of transitions and thereby affect the susceptibilities.

It is very important to note that these couplings lead to correlations in many functions of the membranes. Anesthetics and pressure both influence elastic constants and fluctuation lifetimes. This also implies that pressure influences the effect of anesthetics. Similar couplings exist between capacitance, membrane thickness and electrical phenomena. They are all tied to the same thermodynamics, and by necessity a change in one intensive variable will effect the dependence of the membrane state on the change of other variables. Compressibilities are not independent of heat capacity, capacitance and fluctuation lifetime. 

The thermodynamics of membranes brings about a coherent picture in which all phenomena are coupled in a relatively simple framework. The emergence of pulses and channel-like features in membranes, and the coupling to the time scales are an immediate consequence. Such an approach is in strong contrast to the usual attempt to provide separate theories for each phenomenon.



\end{document}